\documentclass[%
aip,jcp,
amsmath,amssymb,
reprint,%
]{revtex4-1}
\usepackage{graphicx}
\usepackage{dcolumn}
\usepackage{bm}
\usepackage[utf8]{inputenc}
\usepackage[T1]{fontenc}
\usepackage{mathptmx}
\usepackage{etoolbox}
\usepackage{booktabs}
\usepackage{xcolor}
\usepackage{tikz}
\usetikzlibrary{calc,positioning,arrows.meta}
\usepackage[colorlinks,bookmarks=false,citecolor=blue,linkcolor=red,urlcolor=blue]{hyperref}
\makeatletter
\def\@email#1#2{%
	\endgroup
	\patchcmd{\titleblock@produce}
	{\frontmatter@RRAPformat}
	{\frontmatter@RRAPformat{\produce@RRAP{*#1\href{mailto:#2}{#2}}}\frontmatter@RRAPformat}
	{}{}
}%
\makeatother
\begin{document}

	\raggedbottom
	\preprint{AIP/123-QED}
	
	\title[]{Isomer effects on neutral-loss dissociation channels of nitrogen-substituted PAH dications\\}
\author{Sumit Srivastav}\thanks{Electronic mail: \href{mailto:sumit.srivastav@unicaen.fr}{sumit.srivastav@unicaen.fr}}
\affiliation{Normandie Univ., ENSICAEN, UNICAEN, CEA, CNRS, CIMAP, 14000 Caen, France}

\author{Sylvain Maclot}
\affiliation{Normandie Univ., ENSICAEN, UNICAEN, CEA, CNRS, CIMAP, 14000 Caen, France}

\author{Alicja Domaracka}
\affiliation{Normandie Univ., ENSICAEN, UNICAEN, CEA, CNRS, CIMAP, 14000 Caen, France}

\author{Sergio D\'{\i}az-Tendero}
\affiliation{Departamento de Qu\'{i}mica, M\'{o}dulo 13, Universidad Aut\'{o}noma de Madrid, 28049 Madrid, Spain}
\affiliation{Condensed Matter Physics Center (IFIMAC), Universidad Aut\'{o}noma de Madrid, 28049 Madrid, Spain}
\affiliation{Institute for Advanced Research in Chemical Sciences (IAdChem), Universidad Aut\'{o}noma de Madrid, 28049 Madrid, Spain}

\author{Patrick Rousseau}\thanks{Electronic mail: \href{mailto:patrick.rousseau@ganil.fr}{patrick.rousseau@unicaen.fr}}
\affiliation{Normandie Univ., ENSICAEN, UNICAEN, CEA, CNRS, CIMAP, 14000 Caen, France}

\date{\today}

\begin{abstract}
We investigate two nitrogen-containing isomers of polycyclic aromatic hydrocarbons (PAHs), quinoline (Q) and isoquinoline (IQ), of composition C$_9$H$_7$N  in collisions with 7~keV O$^+$ and 48~keV O$^{6+}$ projectile ions. Employing ion-ion coincidence mass spectrometry, we determine branching ratios for H-loss, C$_2$H$_2$-loss, and HCN-loss dissociation channels of Q$^{2+}$ and IQ$^{2+}$. The overall contribution of HCN-loss is found to be the dominant decay channel. A comparison with the results of a parallel experiment on naphthalene, the simplest PAH, reveals that HCN-loss in both isomers has a higher propensity than the analogous C$_2$H$_2$-loss of naphthalene. The positional identity of the nitrogen atom in the two isomers mainly manifests in many-body fragmentation of their dications. Potential energy surfaces of Q$^{2+}$ and IQ$^{2+}$ are further computed to explore complete fragmentation mechanisms. Parent dications (Q$^{2+}$ and IQ$^{2+}$) are identified to isomerize via seven-membered ring structures prior to elimination of C$_2$H$_2$ and HCN. While prompt dissociation is the primary pathway, the dominant channel of each neutral-loss class also exhibits delayed fragmentation. 
\end{abstract}

\maketitle

%

\section{Introduction}
Recent investigations have revealed the importance of polycyclic aromatic hydrocarbons (PAHs) and their nitrogenated analogs (PANHs) in fields of combustion chemistry, environmental chemistry and astrochemistry.~\cite{frenklach20, chen98, hrodmarsson25, ricca01} On Earth, PAHs and PANHs are emitted as a side product in incomplete combustion processes and impose an undesirable toxicity to humans and our environments. They are also known to be the precursor of larger molecules resulting in the formation of soot.~\cite{keller2000} In particular, the demonstrated relevance of neutral and ionic PAHs and PANHs molecules across the astrophysical environments, such as interstellar medium,~\cite{tielens08,li20} planetary atmospheres,~\cite{zhao18, nixon24} and meteorites \cite{naraoka23,sephton02} has made them a central focus of theoretical and experimental research in recent years. In fact the widely discussed mid-infrared 6.2~$\mu$m emission bands, observed from a wide variety of sources ranging from objects in our solar system and interstellar medium, are attributed to PAHs.~\cite{peeters21,nixon19} On absorbing ultraviolet (UV) photons, these polycyclic molecules redistribute absorbed energy to their various vibrational modes, and then de-excite by emitting IR radiation. So observed IR radiations are typical characteristics of aromatic CC and CH vibrational modes. PAHs molecules can also contain atoms other than carbon, for example nitrogen which are so called PANHs. However, it has been proposed that PANHs are essential for reproducing the observed blue-shifted 6.2~$\mu$m bands.~\cite{ricca21}

\begin{figure}[!t]
	\centering
	\includegraphics[width=1\linewidth]{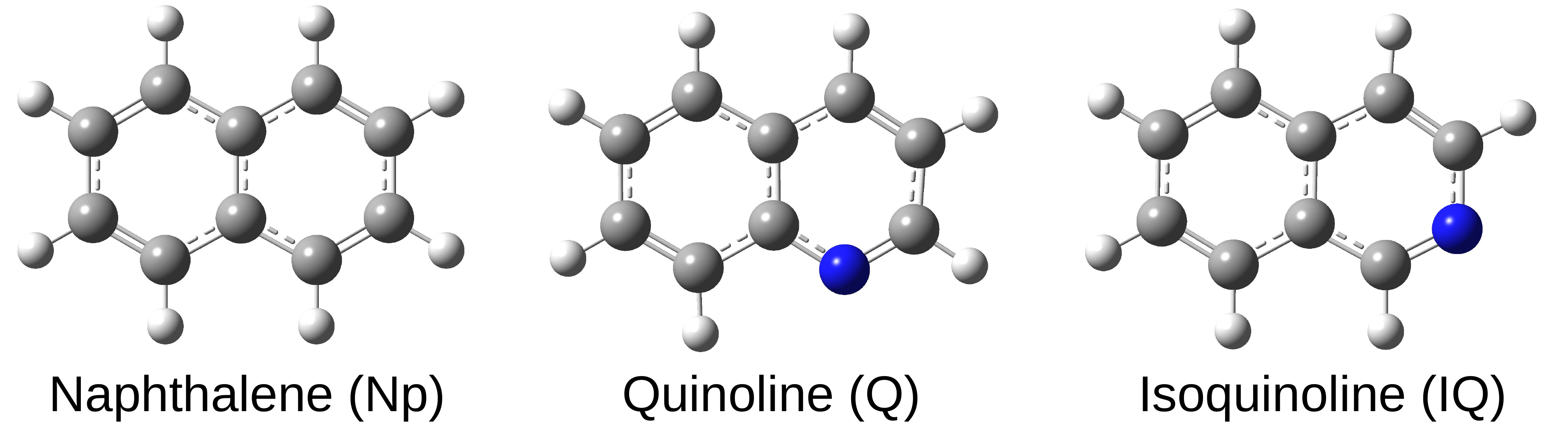}
	\caption{\label{fig:struct}Structures of naphthalene (C$_{10}$H$_8$,~128~au), quinoline and isoquinoline (C$_{9}$H$_7$N,~129~au).}
\end{figure}

In addition, the spectroscopic observations of Cassini-Huygens mission indicate the presence of these molecules in the planetary medium as well, such as Titan.~\cite{nixon24,waite07} The possible mechanisms of N-atom incorporation to the aromatic cyclic molecules in Titan atmosphere are also studied.~\cite{ricca01,vuitton07} It is reported that the complex N-bearing species are precursors for the creation of haze particles and can condense in the Titan's lower stratosphere.~\cite{lebonnois02,wilson04} Moreover, a dense atmosphere and an anoxic environment make Titan a rich source of complex organic reactions leading to the formation of large molecules by associative process. In addition to associative process, it is suggested that the dissociative chemistry induced by different ionizing radiation can also be a prominent source of smaller neutral and charged hydrocarbons and hydrocarbon-nitriles. Up to now, there are eight cyanides including HCN/HNC, HC$_3$N, H$_3$C$_3$N, and CH$_3$CN definitively detected in Titan's neutral atmosphere, and the most abundant ionic species observed is HCNH$^+$ ion.~\cite{nixon24,cravens06} Insights into PANHs excitation, relaxation, and fragmentation are critical to elucidate their formation mechanisms, prevalence, and chemical behavior in planetary atmospheres. Two isomers of C$_9$H$_7$N composition (Fig.\ \ref{fig:struct}), quinoline (Q) and isoquinoline (IQ) are the smallest PANHs molecules postulated as one of the precursors for the observed hydrocarbon-nitriles species.~\cite{nixon24,hrodmarsson25} In view of the significance of these polycyclic molecules, VUV photodissociation studies of Q$^+$ and IQ$^+$ monocations have been reported in recent years.~\cite{bouwman15,kadhane22,arun23,leach18}

In the Titan's atmosphere, keV~ion-collision chemistry can have a significant role in its chemical composition. The singly charged H$^+$ and O$^+$ are the primary available ions which originate from the Saturn's magnetosphere and can penetrate into Titan's atmosphere triggering the local chemistry in its environment.~\cite{cravens08} The keV energy ions are efficient to impart strong perturbation on molecules and may lead to the formation of various molecular multications. Therefore, the formed molecular cations can significantly produce a wide range of ionic and neutral species via their various dissociative pathways. In the last decade, PAHs have been extensively studied under various source of ionizing radiations.~\cite{west12,chen15,reitsma14,gatchell15,chacko22,lozano25,najeeb17} A number of experimental investigations are reported mainly focused  on the fragmentation of simplest PAHs, naphthalene (Np) and its isomer azulene cations produced under ion impacts.~\cite{reitsma13,mishra13,vinitha18} Given the fact that PANHs could be a potential source of various identified small hydrocarbon-nitrile species in the astronomical media, however, there are limited experimental reports available for this class of molecules. 

In this work, we present a systematic experimental study based on the dissociation of the simplest PANHs, quinoline and isoquinoline (Fig.~\ref{fig:struct}), dications produced by keV ion impact. In parallel, we examine naphthalene, the simplest PAH, under the same set of projectiles to assess the influence of a single nitrogen substituent on the dissociation pathways of the analogous PANHs cations. To accomplish this, we irradiate neutral gas-phase naphthalene, quinoline, and isoquinoline with 7~keV O$^{+}$, and to probe a different excitation regime and enable comparison, we have also used 48~keV O$^{6+}$ ion. By using an ion-ion coincidence measurement scheme, we identify various dissociative pathways of Q$^{2+}$ and IQ$^{2+}$ and determine their branching ratios (BRs) for both projectiles. The prominent fragmentation channels of the dications predominantly involve the loss of hydrogen, C$_2$H$_2$, and HCN from Q$^{2+}$ and IQ$^{2+}$. Despite the structural difference arising from the position of the nitrogen atom, the total mass spectra of the two isomers, and the BRs of the aforementioned channels for the monocations Q$^{+}$ and IQ$^{+}$ are nearly identical, a result consistent with earlier VUV studies.\cite{bouwman15,arun23} However, for the dications, the BRs particularly for the HCN-loss channels differ between the two PANHs isomers. Further, we perform the molecular-dynamics (MD) simulations and quantum-chemical calculations to explore the potential-energy surfaces (PES) of the relevant fragmentation channels of Q$^{2+}$ and IQ$^{2+}$.

\section{Experimental details}
The present experiment has been carried out at the ARIBE facility, the low energy ion beam facility at GANIL, Caen, France.~\cite{bernigaud08} The beam of O$^{+}$ and O$^{6+}$ ions are delivered by an Electron Cyclotron Resonance (ECR) ion source. The ion beam is pulsed at a repetition rate of 4~kHz with a width of 500~ns. An effusive jet of neutral target molecules is produced by evaporation of a commercially procured high purity ($>$98\%) powder from Sigma-Aldrich. All three molecules (Np, Q and IQ) have sufficient vapor pressure at room temperature to provide a stable target density without any need of sample heating. The ion and molecular beams interact in a crossed-beam configuration in a linear time-of-flight (TOF) mass spectrometer capable of multi-hit coincidence measurements. The extraction, acceleration, and drift regions of the TOF spectrometer have typical lengths of 110~mm, 30~mm and 1000~mm, respectively. The more details of experimental method and setup can be found elsewhere.~\cite{chandezon94} We record the TOF spectra of the produced fragment ions in event-by-event mode, allowing us to determine the initial charge state of the molecular system and to measure correlations between the charged fragments produced in a single ion–molecule collision. The low ion beam current, of the order of 10--15~pA, and a low event rate ($\sim$50--60~Hz) are maintained to minimize false coincidences. With the help of TOF correlation map, one can identify the particular dissociation channel resulted from a dication fragmentation. 

\section{Computational details}
Quantum chemistry calculations were carried out using the Gaussian16 package~\cite{gaussian16} and in the framework of the Density Functional Theory (DFT). We used the B3LYP~\cite{vosko80,lee88,becke93,stephens94} functional in combination with the 6-31G(d,p) basis set,~\cite{hehre72,hariharan73} a split-valence double-zeta basis set which includes polarization functions polarization on all atoms. Molecular dynamics simulations were carried out using the Atom Centered Density Matrix Propagation (ADMP) method,~\cite{iyengar01,schlegel01,schlegel02} where the nuclei move classically in the potential energy surfaces of the doubly ionized system computed with DFT; both singlet and triplet spin multiplicities were considered. In the trajectories we set a time step of $\Delta t =0.5$~fs and a fictitious electron mass of $\mu=0.1$~amu; in addition, the trajectories were propagated using converged self-consistent field results at each point of the dynamics, thus ensuring adiabaticity. We assumed that the initial geometry of the doubly charged (iso)quinoline is that of the neutral molecule, i.e., a sudden ionization by the collision with the ion, mimicking the experimental conditions. We propagated 200 trajectories up to $t_{max}=500$~fs for each molecule and spin multiplicity; in all cases we assumed an internal excitation energy of $E_{exc}=20$~eV randomly redistributed among the nuclear degrees of freedom in each trajectory. Statistics over all trajectories were subsequently computed to identify the dominant fragmentation channels. Then, we performed an exhaustive exploration of the potential energy surface of the doubly ionized molecules. The relevant points located on the potential energy surfaces, minima and transition states, were characterized by computing the harmonic vibrational frequencies. The number of imaginary frequencies indicates the nature of the point: for minima, all frequencies are real, while transition states exhibit one imaginary frequency, indicating a first-order saddle point on the potential energy surface. This imaginary frequency represents the motion along the reaction coordinate, where atomic rearrangement occurs, bonds cleave and/or formation. We checked the validity of the computed transition states by an intrinsic reaction coordinate (IRC) calculation. An IRC traces the minimum energy path from the transition state, thus confirming that it connects to the correct reactants and products on the potential energy surface. Combining molecular dynamics simulations with exploration of potential energy surfaces is a computational strategy that has been successfully used in the past to infer the evolution and behavior of complex molecular systems under ionizing radiation, as well as to characterize the mechanisms and reaction pathways behind the identification of selected fragments in mass spectrometry; see for example.~\cite{ maclot13, kling19, rousseau20, barreiro21, licht23, mishra24}
\section{Results and discussion}
\begin{figure*}[!t]
	\centering
	\includegraphics[height=0.6\textheight, keepaspectratio]{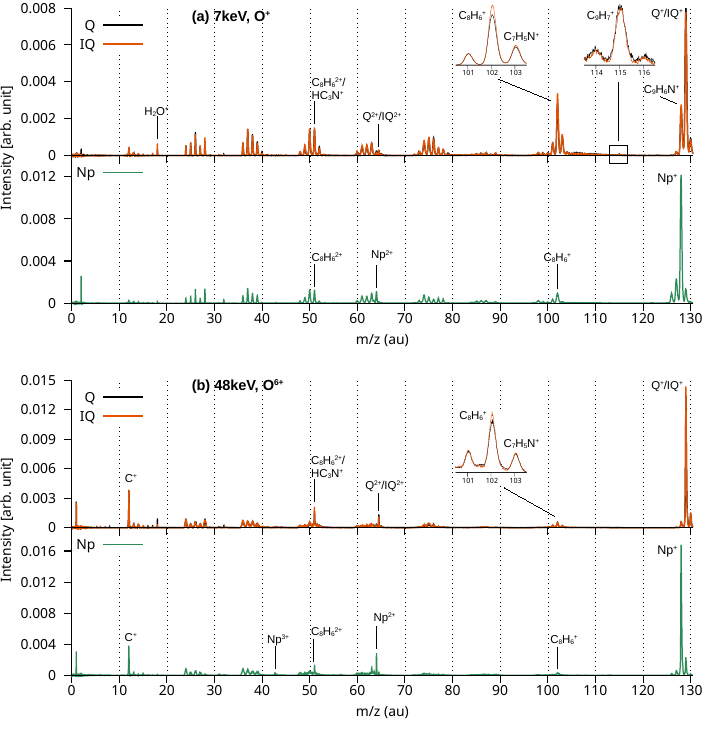}
	\caption{\label{fig:mass1}Total time-of-flight (TOF) mass spectra of cationic products observed in the interaction of neutral target molecules quinoline, isoquinoline and naphthalene with (a) 7~keV O$^+$ and (b) 48~keV O$^{6+}$, projectiles. Each spectrum is normalized by its total ion yield. The water contribution arises from the background. Insets show magnified views of the indicated regions.}
\end{figure*}

\subsection{Mass spectrometry} 
We have performed ion--molecule collision experiments using 7~keV O$^+$ and 48~keV O$^{6+}$ projectile ions on the gaseous molecular targets naphthalene, quinoline and isoquinoline. In ion-molecule collisions, the ionization mechanisms of molecule depend on the charge state of the projectile ion at its given velocity. At keV energy (such that $velocity<$~1~a.u.), highly charged O$^{6+}$ ion mainly ionizes the target molecule via electron capture processes that occurs at larger distances.~\cite{simcic10,srivastav22} In such events, a small amount of energy is transferred to the target, resulting in the formation of rather cold molecular cations.~\cite{rousseau12} On contrary, for a singly charged ion such as O$^{+}$, multiple ionization of the target occurs at lower impact parameters or via penetrating collisions leading to the formation of relatively hotter molecules. In our earlier reports, we have estimated the internal energy distribution of molecules in collision with keV energy singly and multiply charged projectiles.~\cite{maclot16,ewa21}

Fig.\ \ref{fig:mass1}(a) and \ref{fig:mass1}(b) show the total TOF mass spectra for both O$^+$ and O$^{6+}$ ions, each normalized individually to its total yield. Spectra for quinoline and isoquinoline are plotted together (black and orange), and for naphthalene (green), spectrum is shown separately. In both projectile cases, the dominant peak at $m/z=$~129 is the parent ion Q$^+$/IQ$^+$, and it is at $m/z=$~128 for Np$^+$. Peaks beyond the parent mass arise from the $^{13}$C isotope contribution. The doubly charged parent dications Q$^{2+}$/IQ$^{2+}$ ($m/z=$~64.5) and Np$^{2+}$ ($m/z=$~64) are also observed, which indicates that a fraction of these dications are stable up to the $\mu$s time scale within our experimental detection time window. Although the parent dications have smaller yields, Q$^{2+}$/IQ$^{2+}$ are relatively weaker than Np$^{2+}$  for both projectiles, inferring that naphthalene can stabilize positive charge more effectively than the analogous PANHs isomers. All peaks other than the parent cations are fragments emanating in various dissociation pathways of the molecular cations. 

\subsubsection{O$^+$ impact} 
For PAHs and PANHs, the loss of H and loss of C$_2$H$_2$/HCN from a singly charged parent are most commonly discussed statistical dissociation channels in the literature.~\cite{bouwman15,west12,karthick22} Here, for both PANHs isomers (see~Fig.\ \ref{fig:mass1}(a)), the peaks at $m/z=$~102~(C$_8$H$_6^+$) and 128~(C$_9$H$_6$N$^+$) are dominant fragments, which are attributed to the loss of neutral HCN and H from Q$^{+}$/IQ$^{+}$, respectively, and $m/z=$~103~(C$_7$H$_5$N$^+$) is due to the C$_2$H$_2$ loss. From the peak widths and the ion-ion correlation map (discussed later), we can infer that C$_8$H$_6^+$, C$_9$H$_6$N$^+$ and C$_7$H$_5$N$^+$ in the mass spectra have negligible contributions from the decay of Q$^{2+}$/IQ$^{2+}$ dications. A small peak at $m/z=$~127 is due to  loss of 2H/H$_2$. Further, the peak at $m/z=$~51 has two possible cationic assignments: C$_8$H$_6^{2+}$ or HC$_3$N$^+$. The former is the result of HCN emission from Q$^{2+}$/IQ$^{2+}$ and the later is due to the emission of C$_6$H$_6$~(C$_6$H$_6^{+}$) from Q$^{+}$/IQ$^{+}$~(Q$^{2+}$/IQ$^{2+}$). Several additional fragments ranging from lower mass to upward having multiple C atoms with varying intensities can be identified. 

Despite the structural difference due to different position of nitrogen in quinoline vs isoquinoline, (Fig.\ \ref{fig:struct}), their mass spectra and relative yields of primary neutral-loss channels of monocations, Q$^{+}$ and IQ$^{+}$, are strikingly similar. A likewise observation is also reported in a VUV photodissociation study, and it was credited to the fact that both monocations Q$^{+}$ and IQ$^{+}$ isomerize to a common structure which brings different isomers to a common ground before their further dissociation.~\cite{arun23} A small difference in relative intensities of HCN and C$_2$H$_2$ loss channels (Fig.\ \ref{fig:mass1}(a) inset) could stem from the $\sim$0.1~eV difference in ionization energies, with isoquinoline dissociating slightly more abundantly than quinoline.

For naphthalene, H loss from Np$^+$ is the most prominent channel, with 2H/H$_2$ loss also enhanced relative to PANHs (Fig.\ \ref{fig:mass1}(a)~green plot). The $m/z=$~102 (C$_8$H$_6^+$) and $m/z=$~51~(C$_8$H$_6^{2+}$) are primarily due to the emission of C$_2$H$_2$ from Np$^+$ and Np$^{2+}$, respectively, analogous to HCN loss channels of PANHs. As can be seen, compared with (iso)quinoline, the C$_8$H$_6^+$ and C$_8$H$_6^{2+}$ yields are substantially suppressed for naphthalene, consistent with HCN loss being favoured in PANHs over corresponding the C$_2$H$_2$ loss of PAHs. Likewise, the mass spectrum for naphthalene has also broad spectral features of various cationic collision products. However, the overall fragments yield in case of (iso)quinoline is higher than that of naphthalene, which points towards the lower stability of the PANHs under ion impact.

Another key feature that is unique to O$^+$ collision is a peak at $m/z=$~115 (Fig.\ \ref{fig:mass1}(a) inset), which has not been reported in earlier studies. It corresponds to the 14~a.u.~loss from  Q$^{+}$/IQ$^{+}$. We speculate this fragment to be C$_9$H$_7^{+}$ formed through a non-statistical nitrogen knockout process in close collision events between O$^+$ and molecules. Moreover, its absence in the case of naphthalene makes our speculation more affirmative. Stockett \textit{et al.,} reported a detailed investigation of non-statistical knockout processes for anthracene, acridine and phenazine in collisions with helium atoms at 100~eV center-of-mass energy.~\cite{stockett14} The C-atom knockout was found to be energetically less favorable than N-atom knockout.  For naphthalene and for both PANHs, we also do not observe the C loss, which may indicate the instability of C$_8$H$_7$N$^{+}$. In recent investigations, the proposed C$_9$H$_7^{+}$ motif is also identified as an important species, postulated as the 1-indenyl structure ($\mathrm{C_9H_7^{\cdot}}$/$\mathrm{C_9H_7^{+}}$), a fundamental molecular building block to complex PAHs in the interstellar medium.~\cite{yang23,bull25}

\subsubsection{O$^{6+}$ impact}
Fig.\ \ref{fig:mass1}(b) shows the mass spectra for the other projectile O$^{6+}$. It is important to note that the mass distributions obtained under O$^{6+}$ impact are significantly different from those observed for O$^+$ impact. This is a manifestation of different energy deposition to the target molecule in interaction with different projectiles, as discussed earlier. Here, the H and C$_2$H$_2$/HCN neutral-loss dissociation channels of the monocations (Np$^{+}$, Q$^{+}$ and IQ$^{+}$) are scarcely populated, resulting in the weak presence of higher $m/z$ fragments in the mass spectra. Thus, the change in the internal energy of the produced molecular cations appears to be very crucial to the propensity of various neutral-loss channels. 
Moreover, the multiply ionized parent molecular cations are more pronounced under O$^{6+}$ impact. For naphthalene, we even observe Np$^{3+}$ (although weak), whereas the trication is absent for PANHs. This difference reaffirms the better charge delocalization and stabilization capabilities of naphthalene relative to its PANHs counterparts, which enables naphthalene to accommodate extra charge without dissociation.
\begin{figure*}[!t]
	\centering
	\includegraphics[width=1\textwidth]{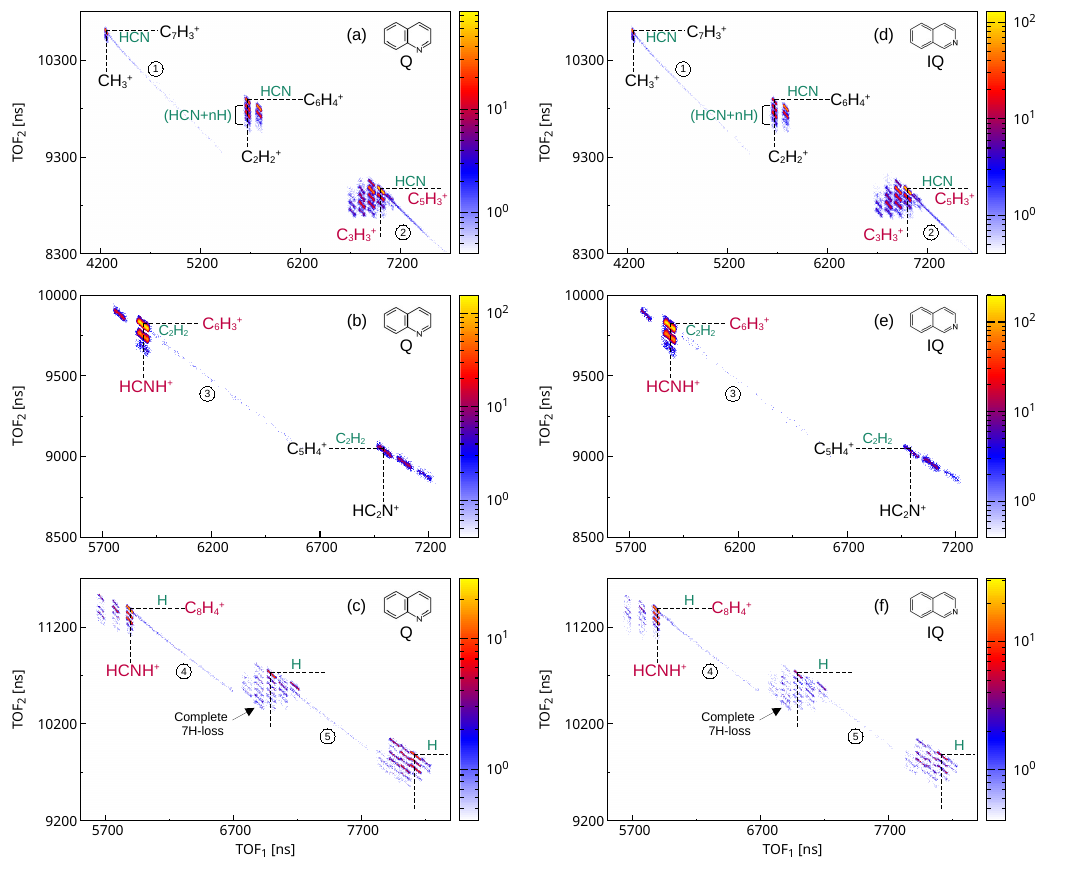}
	\caption{\label{fig:coinc}Selected regions from the coincidence map, where two fragment ions are detected in coincidence, corresponding to different neutral-loss dissociation channels of the quinoline dication (a)-(c) and isoquinoline dication (d)-(f) produced under 48~keV~O$^{6+}$ impact. For reference, some of the ion pairs are labeled with the associated neutral-loss (green), and the dominant neutral-loss channel in each category is highlighted in red. Numbers in circles mark diagonal tail features arising from delayed fragmentation. Intensities are plotted on a logarithmic scale.}
\end{figure*}

Here, for all three molecules, the peak at $m/z=$~51 is relatively sharper, indicating that its primary source is neutral-loss from parent dications. For PANHs isomers, it can arise from the loss of HCN from  Q$^{2+}$/IQ$^{2+}$ and loss of C$_6$H$_6$ from Q$^{+}$/IQ$^{+}$ yielding to C$_8$H$_6^{2+}$ and HC$_3$N$^+$ fragments, respectively, while for naphthalene, $m/z=$~51 is mainly due to the loss of C$_2$H$_2$ from Np$^{2+}$.  Moreover, C$^+$ ($m/z=$~12) attains the highest relative yield in the fragment ion distribution for both class of molecules. The small fragments primarily originate from many-body fragmentation channels of molecular multications. This is typical of charge-driven fragmentation processes, where multiply ionized molecules dissociate by coulombic repulsion into small fragments.~\cite{rousseau12} Interestingly, unlike O$^+$ impact, the $m/z=$~115 peak does not appear in O$^{6+}$ collisions, where single ionization via electron capture at larger impact parameter dominates, making the knockout process less probable.

Beyond neutral-loss from the parent cations, several other features may result from either true two-body breakup or multistep fragmentation of multiply ionized parents with neutral emissions. Later type of fragmentation is the primary subject of our forthcoming discussions. To ascertain such pathways, we turn to adopt ion-ion coincidence measurements. 
    
\subsection{Ion pair correlation}
In this section, we focus on the cationic two-body fragmentation channels that accompany the emission of neutrals from the parent dications, as shown in the following equation:
\begin{eqnarray}
	\mathrm{Np^{2+}/Q^{2+}/IQ^{2+}} &\rightarrow & \mathrm{F_1^+} + \mathrm{F_2^+} +\mathrm{N} \label{channel}
\end{eqnarray}
\begin{table*}[!t]
	\setlength{\belowcaptionskip}{12pt}
	\caption{Branching ratios (BRs) of various neutral-loss channels for both projectiles are given (in \%): (a) HCN-loss and (b) C$_2$H$_2$-loss channels for the PANHs dications, Q$^{2+}$ and IQ$^{2+}$; (c) C$_2$H$_2$-loss channels of PAH dication Np$^{2+}$. BRs are normalized to the total counts of the corresponding full coincidence map. Error bar in BR is approximately~2–3\% of the value.}
	\label{tab:br}
	\centering
	\begingroup
	\footnotesize
	\setlength{\tabcolsep}{4pt}        
	\renewcommand{\arraystretch}{1.05}
	\begin{minipage}[t]{0.35\linewidth}     
		\centering \vspace{0pt}
		\begin{tabular}{@{}l@{\hspace{0.9em}}cc@{\hspace{1.1em}}cc@{}}
			\makebox[0pt][l]{(a)}%
			& \multicolumn{2}{c}{$\mathrm{O}^{+}$} & \multicolumn{2}{c}{~~~$\mathrm{O}^{6+}$} \\
			\specialrule{0.8pt}{0pt}{0pt}
			Ion pair & Q$^{2+}$ & IQ$^{2+}$ & Q$^{2+}$ & IQ$^{2+}$ \\
			CH$_3^{+}$/C$_7$H$_3^{+}$        & 0.57 & 0.93 & 0.70 & 1.08 \\
			C$_2$H$_2^{+}$/C$_6$H$_4^{+}$    & 0.86 & 0.50 & 0.86 & 0.53 \\
			C$_2$H$_3^{+}$/C$_6$H$_3^{+}$     & 1.11 & 1.53 & 1.17 & 1.65 \\
			C$_3$H$_2^{+}$/C$_5$H$_4^{+}$    & 0.55 & 0.38 & 0.38 & 0.27 \\
			C$_3$H$_3^{+}$/C$_5$H$_3^{+}$    & 2.94 & 3.89 & 2.79 & 3.99 \\
			C$_3$H$_4^{+}$/C$_5$H$_2^{+}$    & 0.27 & 0.45 & 0.10 & 0.13 \\
			\addlinespace[0.25em]
			\specialrule{0.5pt}{0pt}{0pt}    
			\addlinespace[0.25em]
			Pure HCN-loss                    & 6.30 & 7.68 & 6.00 & 7.65 \\
			(HCN+nH)-loss                    & 16.62 & 19.57 & 8.33 & 9.96 \\
			\addlinespace[0.25em]
			\specialrule{0.5pt}{0pt}{0pt}    
			\addlinespace[0.25em]
			HCN-loss               & 22.9 & 27.3 & 14.3 & 17.6 \\
			\specialrule{0.7pt}{0pt}{0pt}
		\end{tabular}
	\end{minipage}
	\begin{minipage}[t]{0.3\linewidth}     
		\centering \vspace{0pt}
		\begin{tabular}{@{}l@{\hspace{0.9em}}cc@{\hspace{1.1em}}cc@{}}
			\makebox[0pt][l]{(b)}%
			& \multicolumn{2}{c}{$\mathrm{O}^{+}$} & \multicolumn{2}{c}{~~~$\mathrm{O}^{6+}$} \\
			\specialrule{0.8pt}{0pt}{0pt}
			Ion pair & Q$^{2+}$ & IQ$^{2+}$ & Q$^{2+}$ & IQ$^{2+}$ \\
			HCN$^{+}$/C$_6$H$_4^{+}$         & 0.25 & 0.14 & 0.37 & 0.20 \\
			HCNH$^{+}$/C$_6$H$_3^{+}$        & 3.07 & 3.66 & 3.24 & 4.12 \\
			HC$_2$N$^{+}$/C$_5$H$_4^{+}$     & 0.37 & 0.22 & 0.40 & 0.22 \\
			H$_2$C$_2$N$^{+}$/C$_5$H$_3^{+}$ & 0.36 & 0.42 & 0.33 & 0.37 \\
			H$_3$C$_2$N$^{+}$/C$_5$H$_2^{+}$ & 0.12 & --   & 0.12 & --   \\
			\addlinespace[0.25em]
			\specialrule{0.5pt}{0pt}{0pt}
			\addlinespace[0.25em]
			Pure C$_2$H$_2$-loss              & 4.17 & 4.44 & 4.46 & 4.91 \\
			(C$_2$H$_2$+nH)-loss              & 2.37 & 3.17 & 1.70 & 2.40 \\
			\addlinespace[0.25em]
			\specialrule{0.5pt}{0pt}{0pt}
			\addlinespace[0.25em]
			C$_2$H$_2$-loss            & 6.5  & 7.6  & 6.2  & 7.3  \\
			\specialrule{0.8pt}{0pt}{0pt}
		\end{tabular}
	\end{minipage}
	\begin{minipage}[t]{0.26\linewidth}     
		\centering \vspace{0pt}
		\begin{tabular}{@{}l@{\hspace{0.9em}}cc@{\hspace{1.1em}}cc@{}}
			\makebox[0pt][l]{(c)}%
			& \multicolumn{1}{c@{\hspace{1.2em}}}{$\mathrm{O}^{+}$}
			& \multicolumn{1}{c@{}}{$\mathrm{O}^{6+}$} \\
			\specialrule{0.8pt}{0pt}{0pt}
			Ion pair & Np$^{2+}$ & Np$^{2+}$ \\
			CH$_3^{+}$/C$_7$H$_3^{+}$          & 0.22 & 0.37 \\
			C$_2$H$_2^{+}$/C$_6$H$_4^{+}$      & 0.30 & 0.32 \\
			C$_2$H$_3^{+}$/C$_6$H$_3^{+}$      & 1.11 & 1.30 \\
			C$_3$H$_3^{+}$/C$_5$H$_3^{+}$      & 2.17 & 2.64 \\
			\addlinespace[0.25em]
			\specialrule{0.5pt}{0pt}{0pt}
			\addlinespace[0.25em]
			Pure C$_2$H$_2$-loss               & 3.80 & 4.63 \\
			(C$_2$H$_2$+nH)-loss               & 14.31 & 7.64 \\
			\addlinespace[0.25em]
			\specialrule{0.5pt}{0pt}{0pt}
			\addlinespace[0.25em]
			C$_2$H$_2$-loss            & 18.1 & 12.3 \\
			\specialrule{0.8pt}{0pt}{0pt}
		\end{tabular}
	\end{minipage}
	
	\endgroup
\end{table*}
where F$_1^+$ and F$_2^+$ are the fragment ion pair and N is a neutral species. For a given ion-molecule collision system, there can exist a wide range of ion pairs and neutral losses arising from different collision events. Here, we primarily focus on fragmentation channels for those N~=~(C$_2$H$_2$+nH)/(HCN+nH) (n$\ge0$) and N =mH (m$\ge1$). In the forthcoming discussions, they are collectively referred as C$_2$H$_2$-loss/HCN-loss and H-loss channels. By using the two-stop condition (two charged fragments per event) in coincidence measurements, we identify various dissociation channels of the Np$^{2+}$, Q$^{2+}$ and IQ$^{2+}$ dications; given that most fragments are singly charged in our data. In a coincidence map, the TOF of the first detected fragment is plotted against the TOF of the second fragment, and due to the momentum conservation of the two correlated fragments they appear as a distinct island.  
Fig.\ \ref{fig:coinc}(a)–(c) show the representative regions of coincidence map corresponding to HCN-loss, C$_2$H$_2$-loss, and H-loss of Q$^{2+}$, respectively, whereas Fig.\ \ref{fig:coinc}(d)–(f) show the analogous sets for IQ$^{2+}$ for collisions with the same 48~keV~O$^{6+}$ projectile. The same scheme is applied to the other projectile 7~keV~O$^+$, which is not shown here due to the qualitative similarity.   

\begin{figure*}[!ht]
	\centering
	\includegraphics[width=0.9\textwidth]{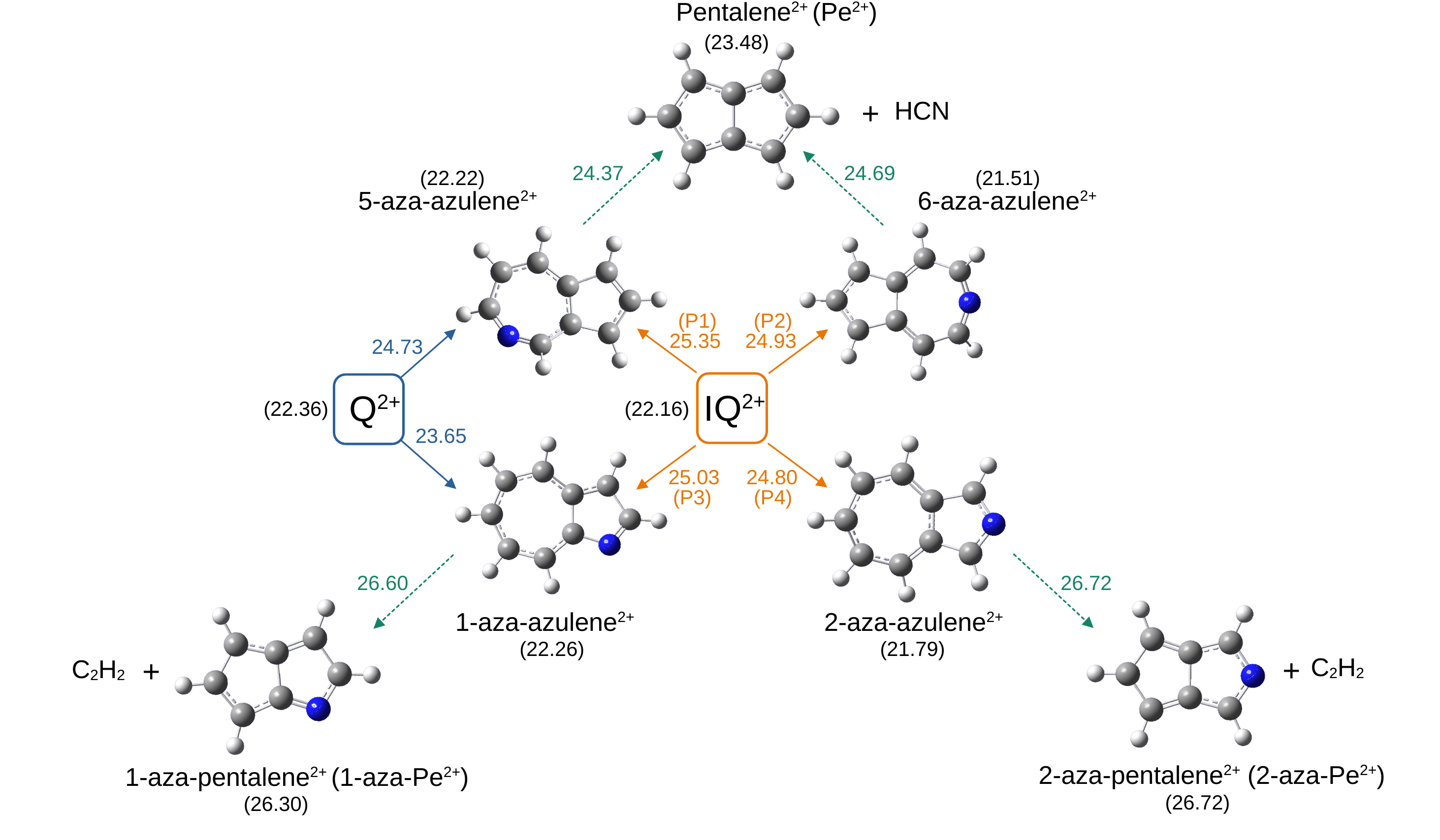}
	\caption{\label{fig:com_path}Scheme summarizing isomerization and subsequent neutral-loss dissociation pathways of the quinoline and isoquinoline dications. Blue and orange arrows connect isomerization structures of Q$^{2+}$ and IQ$^{2+}$, respectively and dotted green arrows indicate the neutral-loss channels. Highest barrier in each step is mentioned along the arrows. Energies are given in eV and are relative to neutral quinoline and isoquinoline.}
	\noindent\rule{\linewidth}{0.6pt} 
\end{figure*}

\subsubsection{HCN-loss channel}
In Fig.\ \ref{fig:coinc}(a) and \ref{fig:coinc}(d), six pure HCN-loss channels (n$=$0) of both isomers dications can be identified, and they are tabulated in Table \ref{tab:br}(a) with their corresponding BRs for both projectiles. The BRs are normalized to the total counts of the full coincidence map (shown in the supplementary material), and the contributions from more-than doubly ionized parents are removed. While assigning the molecular formula to the observed fragment ion of a given mass, literature reports were also taken into account where needed.~\cite{leach18,reitsma13,leach89} We find that the C$_3$H$_3^{+}$/C$_5$H$_3^{+}$ is the dominant fragmentation channel associated with the loss of HCN across both molecules and projectiles. The next most intense channel is C$_2$H$_3^{+}$/C$_6$H$_3^{+}$ ($>$1\% BR), although this pair may include some contribution from HCN$^{+}$/C$_6$H$_3^{+}$ with C$_2$H$_3$ neutral emission. However, with the known HCN-loss propensity in PANHs, we assign this pair as C$_2$H$_3^{+}$/C$_6$H$_3^{+}$. Also the slope of islands suggests that the main dissociation mechanism is neutral emission first, followed by the fission of the remaining  motif C$_8$H$_6^{2+}$ into the ion pair. Such information based on the slope of islands has been widely used in various studies.~\cite{reitsma13,galindo24,eland91} Moreover, two of the HCN-loss channels exhibit delayed fragmentation, which appear as long diagonal tails in the coincidence map (tails \textcircled{1} and \textcircled{2} in Fig.\ \ref{fig:coinc}(a)). The details regarding delayed fragmentation are discussed in a separate section.

Notably, in addition to pure HCN-loss there are ion pairs corresponding to dissociation in which additional hydrogen atoms are emitted; one such instance is displayed in Fig.\ \ref{fig:coinc}(a) and \ref{fig:coinc}(d) in parentheses (HCN+nH), which are described as sequential emission. The islands are well separated by unit mass of hydrogen, which allows us to determine their individual BRs. Although individually none of them are as intense as the strongest pure HCN-loss channel, altogether sequential emission of hydrogen contributes significantly. The multiple hydrogen-loss can proceed as successive emission of hydrogen atoms or as molecular H$_2$, depending on the internal excitation of the parent dication. The sequential emission of HCN and H (n=1) can also be viewed as the loss of intact HCNH. Indeed, in prior VUV photodissociation studies, the sequential loss of HCN and H from monocation Q$^+$ was mainly observed to exist,~\cite{leach18,kadhane22} and here dication fragmentation may be energetically more favorable for such sequential emission. Table \ref{tab:br}(a) shows that the BR for pure HCN-loss differs by around 1.5\% between quinoline and isoquinoline for both projectiles. Adding the contributions from pure and sequential emission, the differences in BRs of total HCN-loss between quinoline and isoquinoline are 4.4\%  and 3.3\% for O$^+$ and O$^{6+}$ projectiles, respectively. Moreover, in every case the sequential BR collectively exceeds the total pure HCN-loss BR. Particularly, for O$^+$ it is approximately 2.6 times larger for both isomers, consistent with the greater internal heating in O$^+$ collisions enhancing sequential hydrogen emission. Altogether, Table \ref{tab:br}(a) summarizes two important points:~(1)~the loss of HCN (either pure or sequential) is consistently higher for isoquinoline than quinoline, and (2)~in internally heated PANHs, abstraction of hydrogens is an important process associated with HCN elimination pathways. 

The loss of neutrals and the fragmentation mechanisms are further studied with the exploration of PES, an overall scheme for a HCN loss from Q$^{2+}$ and IQ$^{2+}$ is shown in Fig.\ \ref{fig:com_path}. We find that both quinoline and isoquinoline isomerize to seven-membered ring structures before the elimination process of HCN. Similar seven-membered ring isomerization mechanism was reported by Dyakov \textit{et al.,} for naphthalene monocation Np$^+$ and in a later study it was also reported for Q$^+$ and IQ$^+$.~\cite{arun23, dyakov06} As can be seen both Q$^{2+}$ and IQ$^{2+}$ isomerize to a common structure 5-aza-azulene$^{2+}$ which further emit HCN with the formation of pentalene$^{2+}$ (Pe$^{2+}$). However, IQ$^{2+}$ exhibits an additional pathway P2 leading to the same product via isomerization to 6-aza-azulene$^{2+}$. 

\begin{figure*}[!ht]
	\centering
	\includegraphics[width=1\linewidth]{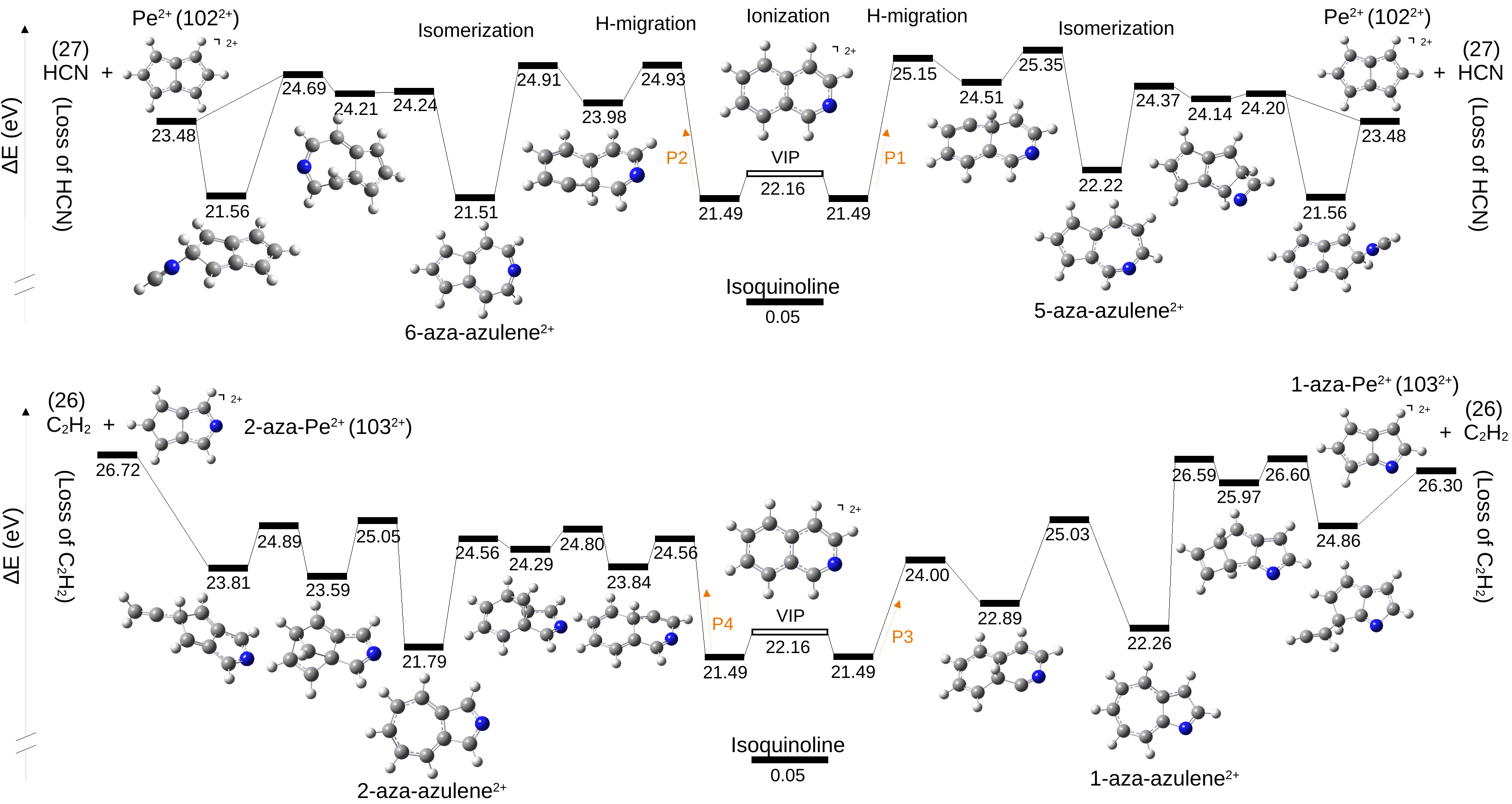}
	\caption{\label{fig:PEC2}Potential energy surfaces (PES) are summarized for a loss of HCN (top) and C$_2$H$_2$ (bottom) from IQ$^{2+}$ for its both possible pathways. }
\end{figure*}

The top panel of Fig.\ \ref{fig:PEC2} describes the complete PES for HCN loss of IQ$^{2+}$ via its both pathways P1 and P2. The energies of all structures and states are in eV and are referenced to the energy of neutral ground state isoquinoline. It is evident that both pathways proceed via hydrogen migration followed by the corresponding isomerization.
Afterward, the elimination of HCN begins via ring deformation and rearrangement of the HCN group, which further proceed to its elimination from the ring, forming Pe$^{2+}$. The highest barriers for isomerization in P1 and P2 are 25.35~eV and 24.93~eV, respectively, which exceed the barriers of their corresponding HCN elimination reactions, 24.37~eV and 24.69~eV. It tells that if the system has sufficient internal energy to cause isomerization, the corresponding HCN-loss reaction from the isomer becomes naturally favorable.

\begin{figure*}[!ht]
	\centering
	\includegraphics[width=0.8\linewidth]{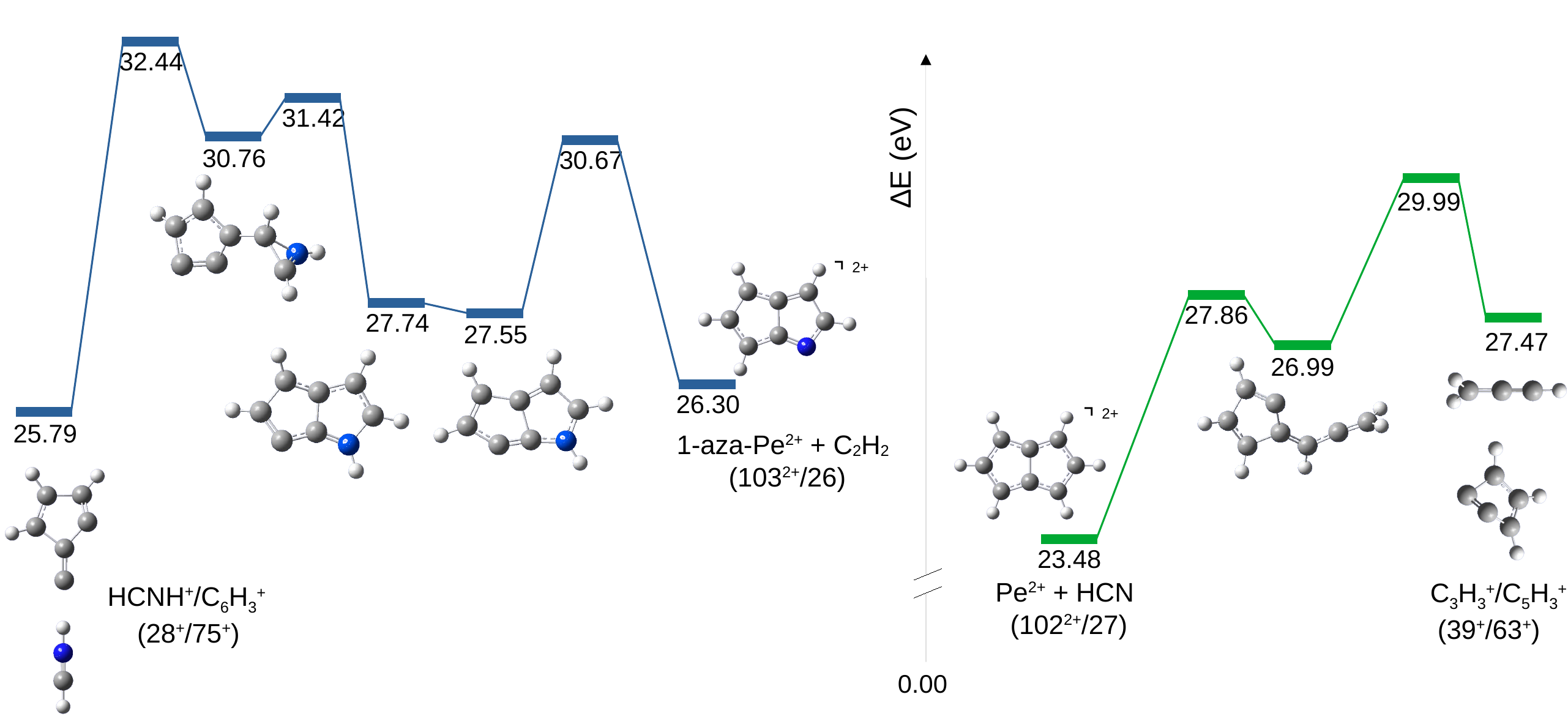}
	\caption{\label{fig:PEC1} Further fragmentation of Pe$^{2+}$ and 1-aza-Pe$^{2+}$ into C$_3$H$_3^{+}$/C$_5$H$_3^+$ (right) and HCNH$^{+}$/C$_6$H$_3^+$ (left) ion pairs, respectively. }
	\noindent\rule{\linewidth}{0.6pt} 
\end{figure*}

Thus, in the mass spectra (Fig.\ \ref{fig:mass1}), the peak at $m/z=$~51 with a possible contribution from C$_8$H$_6^{2+}$ can be attributed to Pe$^{2+}$. The presence of this peak indeed supports its definite stability, however, we find that C$_3$H$_3^{+}$/C$_5$H$_3^+$, following the loss of HCN is the dominant fragmentation channel of both PANHs parent dications (see Table \ref{tab:br}(a)). The PES details of further fragmentation of Pe$^{2+}$ to its most probable C$_3$H$_3^{+}$/C$_5$H$_3^+$ channel are shown in the right branch of Fig.\ \ref{fig:PEC1}. The state at 29.99~eV determines the highest barrier for the reaction. The fragmentation pattern from Pe$^{2+}$ is identical regardless of the starting parent, it is intriguing to see that the BR of C$_3$H$_3^{+}$/C$_5$H$_3^+$ (HCN) is observed to be consistently higher for isoquinoline. This can be attributed to the existence of two possible isomerization pathways P1 and P2 (energetically very similar) to form Pe$^{2+}$ making HCN elimination more probable for isoquinoline than quinoline. 

Likewise, quinoline also reaches the same products via the isomerization and fragmentation processes. The corresponding PES details are given in the supplementary material~(SM). 

\subsubsection{C$_2$H$_2$-loss channel}		
Fig.\ \ref{fig:coinc}(b) and \ref{fig:coinc}(e) present islands corresponding to C$_2$H$_2$-loss channels of Q$^{2+}$ and IQ$^{2+}$, respectively with their BRs summarized in Table \ref{tab:br}(b). For both dications, HCNH$^{+}$/C$_6$H$_3^{+}$ is the strongest fragmentation channel following the emission of C$_2$H$_2$. In fact this channel is observed to be the most intense across all neutral-loss channels, except for IQ$^{2+}$ under O$^{+}$ where the HCN-loss channel C$_3$H$_3^{+}$/C$_5$H$_3^{+}$ (3.89\%) slightly exceeds the corresponding C$_2$H$_2$-loss BR (3.66\%). The contribution from sequential hydrogen emission accompanying C$_2$H$_2$ loss is not as significant as it is in the case of HCN-loss. Moreover, the BRs for pure C$_2$H$_2$-loss are essentially identical for Q and IQ, and their total BRs differ by 1.1\% for both projectiles.

\begin{figure*}[!ht]
	\centering
	\includegraphics[width=0.95\linewidth]{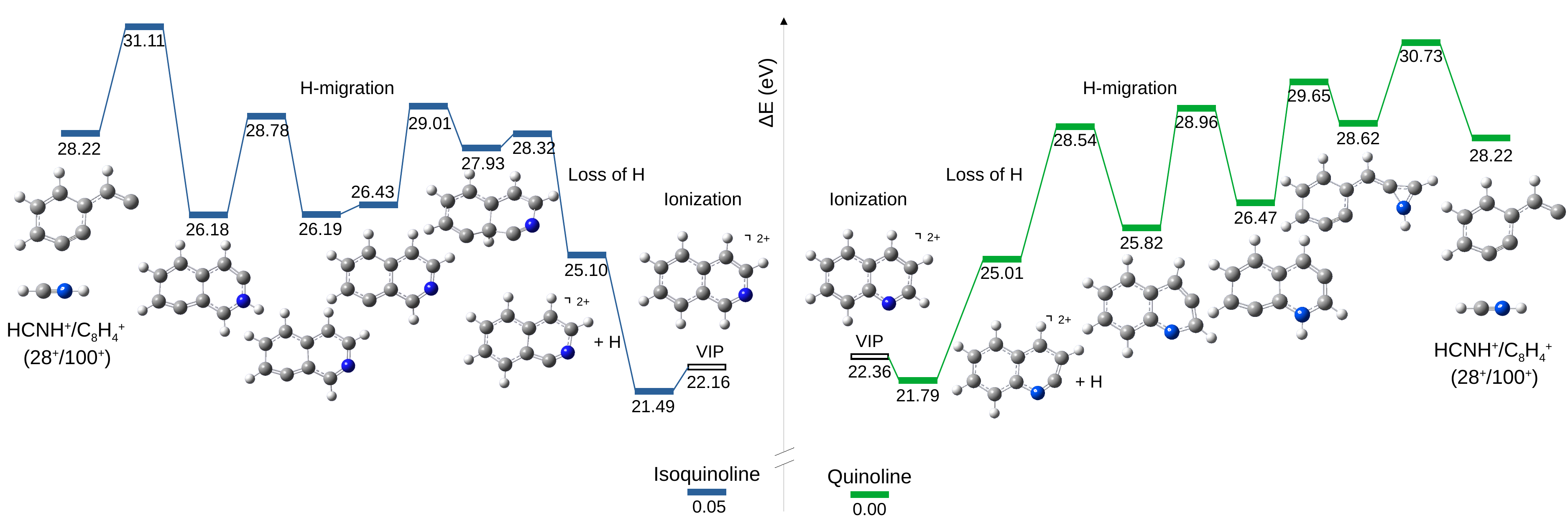}
	\caption{\label{fig:PEC3}The loss of hydrogen atom from Q$^{2+}$ (right) and IQ$^{2+}$ followed by the further fragmentation into HCNH$^+$/C$_8$H$_4^+$ ion pair.}
\end{figure*}
In parallel, the C$_2$H$_2$-loss channels of Np$^{2+}$ produced under the same set of projectiles were also examined. From Table \ref{tab:br}(c), we see that the C$_3$H$_3^{+}$/C$_5$H$_3^{+}$ fragment pair has the highest contribution following the loss of C$_2$H$_2$ from Np$^{2+}$, and the second most intense breakup pair is C$_2$H$_3^{+}$/C$_6$H$_3^{+}$. In a 30~keV He$^{2+}$ collision study of naphthalene, Reitsma \textit{et al.,} demonstrated the same two channels as dominant but in reverse order of their relative intensities (Table 1 of ref \cite{reitsma13}). Notably, C$_2$H$_2$ loss from Np$^{2+}$ is an analogous process to HCN loss from Q$^{2+}$/IQ$^{2+}$, as both processes yield the same isoelectronic C$_8$H$_6^{2+}$ motif before dissociation.  A clear similarity can also be seen in their respective ion pairs (see Table \ref{tab:br}(a) and~\ref{tab:br}(c)). Likewise, the sequential emission channel C$_2$H$_2$+nH has a significant contribution here as well. In fact, in the same report by Reitsma \textit{et al.,} the sequential channel C$_2$H$_2$+2H of Np$^{2+}$ was observed to be equally strong as the dominant pure C$_2$H$_2$ loss channel. 

As in the case of HCN loss, isoquinoline exhibits two pathways, P3 and P4, to emit C$_2$H$_2$ via isomerization to 1-aza-azulene$^{2+}$ and 2-aza-azulene$^{2+}$ structures, respectively~(see~Fig.\ \ref{fig:com_path}). Quinoline again follows one common pathway via 1-aza-azulene$^{2+}$ isomer formation, and the corresponding PES details are provided in the SM. Fig.\ \ref{fig:PEC2} (bottom panel) summarizes the complete PES for C$_2$H$_2$ loss from IQ$^{2+}$ for both pathways. In a similar manner, P3 and P4 follow hydrogen migration before forming 1-aza-azulene$^{2+}$ and 2-aza-azulene$^{2+}$ isomers by crossing their corresponding barriers at 25.03~eV and 24.80~eV, respectively. Further, with deformation of the bigger ring and bond rearrangements, the elimination of C$_2$H$_2$ takes place. It is important to emphasize that we observe the C$_2$H$_2$ elimination from the non-nitrogen containing ring only.

By contrast with HCN loss, the barrier for C$_2$H$_2$ loss is higher than the respective isomerization barrier for both P3 and P4.  Also from the comparison between barrier heights of the two loss channels, we find that the abstraction of C$_2$H$_2$ is more endoergic ($\approx$2~eV) than losing HCN, which can explain the overall less contribution of the former in BRs (see~Table~\ref{tab:br}). This may also account for the less significant difference in C$_2$H$_2$-loss BRs between quinoline and isoquinoline, despite the two possible pathways for isoquinoline.

Further, a complete PES leading to the strongest HCNH$^+$/C$_6$H$_3^+$ channel observed in the fragmentation of 1-aza-Pe$^{2+}$ (a common isomer for Q and IQ) is illustrated in the left branch of Fig.\ \ref{fig:PEC1}, and the additional pathway of IQ$^{2+}$ forming the same ion pair from the fragmentation of 2-aza-Pe$^{2+}$ is provided in the SM. The fragmentation begins with similar ring deformation and finally proceeds to dissociation via a state at 32.44~eV that sets the highest barrier for the reaction. This barrier is higher than that for C$_3$H$_3^+$/C$_5$H$_3^+$~(HCN) by 2.45~eV. Despite this, it is intriguing that the BR of HCNH$^+$/C$_6$H$_3^+$ (C$_2$H$_2$) channel is marginally higher than the BR of C$_3$H$_3^+$/C$_5$H$_3^+$~(HCN) channel, except for isoquinoline under O$^+$ impact. In the only report available of dissociative double photoionization of quinoline, Leach \textit{et al.,} reported a similar trend in BRs of these two fragmentation channels. (Table V of ref\cite{leach89}).

\subsubsection{H-loss channel}
Finally, we present the H-loss dissociation channels of Q$^{2+}$, IQ$^{2+}$ and Np$^{2+}$. For both PANHs isomers, HCNH$^+$/C$_8$H$_4^+$ breakup with the emission of a single hydrogen is found to be the strongest channel (see Fig.\ \ref{fig:coinc}(c) and \ref{fig:coinc}(f)). For naphthalene the corresponding channel is observed to be C$_3$H$_3^+$/C$_7$H$_3^+$ with the loss of 2H (coincidence map is shown in SM). Although the intensities are weak, we also observe complete seven-hydrogen and eight-hydrogen emissions from Q$^{2+}$/IQ$^{2+}$ and Np$^{2+}$, respectively, and the corresponding islands are indicated in the coincidence maps. Several other H-loss channels are clearly identified, and we notice a key tendency that both PANHs lose a single or odd number of hydrogen atoms preferentially in their dicationic fragmentation, whereas the PAH (naphthalene) favours the emission of an even number of hydrogens. Due to several H-loss channels, we have not given a separate table of BRs, instead their combined yields are discussed in the later part of the section.

The complete PES for the strongest HCNH$^+$/C$_8$H$_4^+$ fragmentation channel of C$_9$H$_6$N$^{2+}$ following the emission of a hydrogen from Q$^{2+}$ and IQ$^{2+}$ are shown in Fig.\ \ref{fig:PEC3}. The energy required for a hydrogen atom abstraction from the parent dications is around 25~eV relative to the neutral ground state. We find that H-migration is the primary process leading to this fragmentation channel. The transition states at 30.73~eV and 31.11~eV determine the highest barriers of the reaction for Q$^{2+}$ and IQ$^{2+}$, respectively.  
This small difference in barriers makes H loss slightly more likely for quinoline than for isoquinoline, that is consistent with our experimental observation. Likewise, we expect that the loss of multiple hydrogens and subsequent fragmentation can occur via multiple H-migrations, which may involve higher reaction barriers.
\begin{figure}[!t]
	\centering
	\begingroup
	\begin{minipage}[t]{0.9\linewidth}
		\centering \vspace{0pt}
		\includegraphics[width=\linewidth]{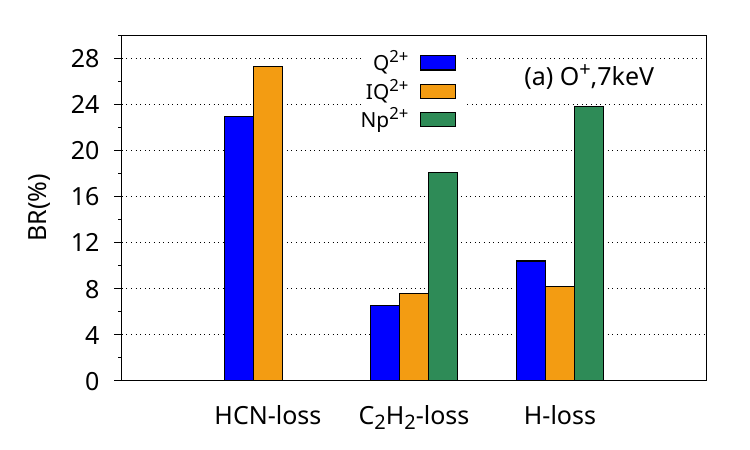}
	\end{minipage}
	\hspace{0.01\linewidth} 
	\begin{minipage}[t]{0.9\linewidth}
		\centering \vspace{0pt}
		\includegraphics[width=\linewidth]{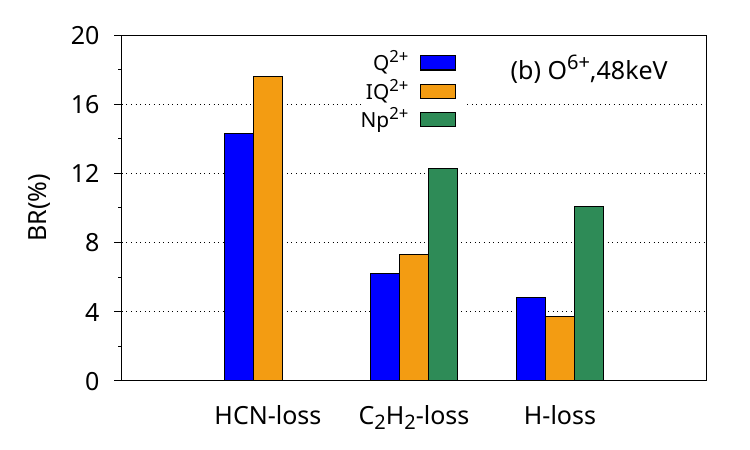}
	\end{minipage}
	\caption{Branching ratios of neutral-loss channels for all three parent dications, shown for (a) 7~keV~O$^{+}$ and (b) 48~keV~ O$^{6+}$ projectiles.  Error bar on the BR is approximately 2-3\% of the value.}
	\label{fig:Bar_BR}
	\endgroup
\end{figure}

For a direct comparison across molecules and projectiles, Fig.\ \ref{fig:Bar_BR} compiles the BRs for each neutral-loss class. It can be seen in Fig.\ \ref{fig:Bar_BR}(a) that under O$^+$ impact, the H-loss channel dominates for Np$^{2+}$ followed by the C$_2$H$_2$-loss. Quinoline and isoquinoline also depict the same trend for these two channels, with H-loss exceeding the C$_2$H$_2$-loss. However, under O$^{6+}$ impact, in Fig.\ \ref{fig:Bar_BR}(b), this order is reversed for all three molecules, highlighting the key role of internal excitation of molecules in their multistep fragmentation. Nevertheless, for the two studied PANHs isomers, it is the HCN-loss channel which always dominates, and its BR is consistently higher for isoquinoline with a contrastingly higher difference. Moreover, it is important to emphasize that all three neutral-loss classes collectively make a significant contribution to the overall dication fragmentation. For O$^+$ impact, they sum to 39.8\%, 43.1\% and 41.9\% for Q$^{2+}$, IQ$^{2+}$ and Np$^{2+}$, respectively, and the corresponding totals for O$^{6+}$ are 25.3\%, 28.6\% and 22.4\%.

In addition to simulations exploring the potential energy surface to identify the mechanisms of the experimentally observed fragmentation channels, molecular dynamics simulations were also carried out. Doubly ionized quinoline and isoquinoline molecules with a certain amount of excitation energy were considered and molecular dynamics simulations were run for 500~fs. In this way, the fragmentation channels that initially occur after ionization and excitation in the collision could be identified. Confirming the experimental results, the loss of the neutral fragments H, HCN, and C$_2$H$_2$ were the most populated breakup channels (see SM).

\subsection{Delayed fragmentation}
As mentioned in the previous section, the coincidence map contains a few islands with diagonal tails that are associated with delayed fragmentation of the metastable dicationic states (see Fig.\ \ref{fig:coinc}). By contrast, the intense part of each island is due to the dominant prompt dissociation. It is interesting to note that the strongest channel of each neutral-loss class across all three molecules exhibits a delayed component. For quinoline and isoquinoline, these channels are C$_3$H$_3^{+}$/C$_5$H$_3^{+}$ (HCN), HCNH$^{+}$/C$_6$H$_3^{+}$ (C$_2$H$_2$) and HCNH$^{+}$/C$_8$H$_4^{+}$ (H). For naphthalene, the respective channels are C$_3$H$_3^{+}$/C$_5$H$_3^{+}$ (C$_2$H$_2$) and C$_3$H$_3^{+}$/C$_7$H$_3^{+}$ (2H).

Furthermore, the lifetime of a metastable precursor can be approximated by performing a time analysis along the diagonal tails.~\cite{field93,galindo24,wei21} From the intensity information along the structure of a tail, the decay constant of the metastable dication can be determined by adopting a fitting method with an exponential formula given by T. A. Field and J. H. Eland.~\cite{field93} Some of the representative graphs and fitting processes are given in the SM. In the past, we have used a similar method to determine the decay constant information of metastable states.~\cite{galindo24}

Under O$^{6+}$ impact, we determine $\tau=622\pm3$~ns and 1830$\pm$185~ns for C$_3$H$_3^{+}$/C$_5$H$_3^{+}$ (HCN) and HCNH$^{+}$/C$_8$H$_4^{+}$ (H) channels of Q$^{2+}$, respectively. For isoquinoline we find the values to be nearly identical to those of quinoline. For both PANHs, the intensities along the tail corresponding to HCNH$^{+}$/C$_6$H$_3^{+}$~(C$_2$H$_2$) are too low to deduce associated $\tau$ values. Similarly for naphthalene, the $\tau$ values are $644\pm6$~ns and $1591\pm148$~ns for C$_3$H$_3^{+}$/C$_5$H$_3^{+}$~(C$_2$H$_2$) and C$_3$H$_3^{+}$/C$_7$H$_3^{+}$ (2H) channels, respectively. The obtained values $\tau=622\pm3$~ns for C$_3$H$_3^{+}$/C$_5$H$_3^{+}$ (HCN) of Q$^{2+}$/IQ$^{2+}$ and $644\pm6$~ns for C$_3$H$_3^{+}$/C$_5$H$_3^{+}$~(C$_2$H$_2$) of Np$^{2+}$ are very close to each other that reflects the resemblance of two decay processes with the same metastable precursor C$_8$H$_6^{2+}$. In  case of O$^+$ impact, the features of delayed fragmentation are broadly similar, except that the observed diagonal tails are approximately 1.5 to 2 times less populated than under O$^{6+}$ impact. This is consistent with a larger prompt dissociation probability at higher internal excitation by O$^{+}$. 

We find another key feature that is unique to C$_2$H$_2$ neutral-loss across all three parent dications under both projectiles. As a representative case, a part of the coincidence map of Q$^{2+}$ fragmentation for the O$^{6+}$ ion is shown in Fig.\ \ref{fig:C2H2}. Here, in addition to the diagonal tails, a vertical tail feature appears in coincidence with the HCNH$^+$ fragment ion. We attribute this feature to dissociation of the metastable state of C$_8$H$_5^{+}$ intermediate that is formed together with HCNH$^+$ in the first step (eqn.~\ref{neut_ms1_a}), followed by delayed emission of C$_2$H$_2$ neutral at a later position in the spectrometer to produce C$_6$H$_3^{+}$~(eqn.~\ref{neut_ms1_b}). The C$_6$H$_3^{+}$ fragment therefore remains in coincidence with HCNH$^+$, creating a vertical tail in the map. For isoquinoline an identical vertical tail is observed, associated with the same delayed fragmentation pathways as in quinoline.

\begin{subequations}\label{neut_ms1}
	\refstepcounter{equation}\label{neut_ms1_a} 
	\refstepcounter{equation}\label{neut_ms1_b} 
\end{subequations}

\begin{figure}[h]
	\centering
	\begin{tikzpicture}[>=Stealth,thick,baseline=(A.base)]
		\def\Main{3.0cm}   
		\def\Drop{0.5cm}   
		\def\Branch{1.8cm} 
		
		\node (A) at (0,0) {$\mathrm{C_9H_7N^{2+}}$};
		\node (B) at (\Main,0) {$\mathrm{C_8H_5^{+}} + \mathrm{HCNH^{+}}$};
		
		\coordinate (Dstart) at ($(B.south west)+(0.3,0)$);
		\coordinate (Dend)   at ($(Dstart)+(0,-\Drop)$);
		
		\node (C) at ($(Dend)+(\Branch,0)$) {$\mathrm{C_6H_3^{+}} + \mathrm{C_2H_2}$};
		
		\draw[-{Stealth[length=3mm]}] (A) -- (B);           
		\draw[densely dotted]         (Dstart) -- (Dend);   
		\draw[-{Stealth[length=3mm]}] (Dend) -- (C.west);   
		
		\node[right=8.8mm] at (B.east) {\textup{(\ref*{neut_ms1_a})}};
		\node[right=2mm] at (C.east) {\textup{(\ref*{neut_ms1_b})}};
	\end{tikzpicture}
\end{figure}

\begin{figure}[!t]
	\centering
	\includegraphics[width=1\linewidth]{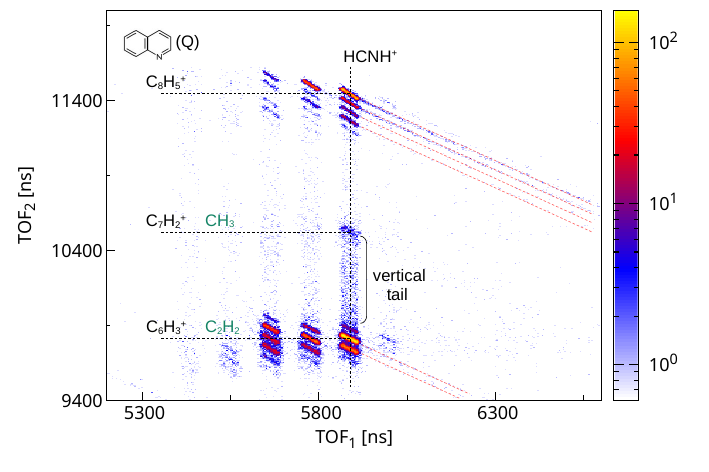}
	\caption{\label{fig:C2H2}A region of the ion-ion coincidence map for quinoline and 48~keV O$^{6+}$ collision system. Diagonal and vertical tails, in coincidence with HCNH$^+$ fragment ion, corresponding to different delayed fragmentation mechanisms. Details are discussed in the text.}
\end{figure}


A similar feature (although weaker) is also observed for naphthalene, and the analogous breakup pathways place the vertical tail in coincidence with C$_2$H$_3^{+}$ (eqn.~\ref{neut_ms2}). To the best of our knowledge, such a delayed fragmentation feature with a vertical tail has not been reported in the fragmentation of  isolated molecules. However, a similar behavior has been described in coincidence measurements of the fragmentation of uracil clusters (Fig.3 of ref~\cite{pal16}). 

\begin{subequations}\label{neut_ms2}
	\refstepcounter{equation}\label{neut_ms2_a} 
	\refstepcounter{equation}\label{neut_ms2_b} 
\end{subequations}

\begin{figure}[h]
	\centering
	\begin{tikzpicture}[>=Stealth,thick,baseline=(A.base)]
		\def\Main{2.7cm}   
		\def\Drop{0.5cm}   
		\def\Branch{1.8cm} 
		
		\node (A) at (0,0) {$\mathrm{C_{10}H_8^{2+}}$};
		\node (B) at (\Main,0) {$\mathrm{C_8H_5^{+}} + \mathrm{C_2H_3^+}$};
		
		\coordinate (Dstart) at ($(B.south west)+(0.3,0)$);
		\coordinate (Dend)   at ($(Dstart)+(0,-\Drop)$);
		
		\node (C) at ($(Dend)+(\Branch,0)$) {$\mathrm{C_6H_3^{+}} + \mathrm{C_2H_2}$};
		
		\draw[-{Stealth[length=3mm]}] (A) -- (B);           
		\draw[densely dotted]         (Dstart) -- (Dend);   
		\draw[-{Stealth[length=3mm]}] (Dend) -- (C.west);   
		
		\node[right=12mm] at (B.east) {\textup{(\ref*{neut_ms2_a})}};
		\node[right=2mm] at (C.east) {\textup{(\ref*{neut_ms2_b})}};
	\end{tikzpicture}
\end{figure}

In Fig.\ \ref{fig:C2H2}, the slope of diagonal tail associated with HCNH$^+$/C$_6$H$_3^{+}$ (C$_2$H$_2$) breakup is approximately -1. It suggests that the emission of C$_2$H$_2$ neutral is followed by the fission of the metastable dication C$_7$H$_5$N$^{2+}$ into the ion pair, whereas the vertical tail corresponds to ion pair formation followed by C$_2$H$_2$ neutral emission. The intense island with slope -1 corresponds to the prompt dissociation process. It is interesting to note that all these different pathways coexist in the same system and involve the emission of C$_2$H$_2$ either in the first step or in the second step of fragmentation. Furthermore, the ions in coincidence with the HCNH$^+$ fragment exhibit several diagonal tails associated with delayed fragmentation of dication metastable states. It suggests that the presence of nitrogen atom in PANHs isomers opens up various dicationic metastable states. Moreover, the HCNH$^+$ cation is one of the highest contributing fragments observed in the fragmentation of Q$^{2+}$ and IQ$^{2+}$ with total abundances of 13\% and 17\% (for both projectiles), respectively. The dominance of neutral HCN and HCNH$^+$ cation in ion-induced dissociative processes of PANHs indeed can be one of the explanation for the abundant detection of these species in Titan's atmosphere. 

\section*{Conclusion}  
The cationic two-body fragmentation channels of two PANHs isomer dications Q$^{2+}$ and IQ$^{2+}$, accompanying with their primary neutral-loss channels H-loss, C$_2$H$_2$-loss, and HCN-loss have been studied in collisions with 7~keV O$^{+}$ and 48~keV O$^{6+}$ projectile ions. Despite the structural difference arising from the position of the nitrogen atom, the overall mass spectra of the two isomers were observed to be generally very similar. To assess the effect of a nitrogen substituent, we also examined the analogous pure-hydrocarbon PAH, naphthalene. Branching ratios for the various dissociation channels were obtained using ion-ion coincidence measurements. We found that H-loss and C$_2$H$_2$/HCN neutral-loss channels collectively contribute a significant fraction of the total dication fragmentation yields of Q$^{2+}$, IQ$^{2+}$, and Np$^{2+}$. The relative intensities of total H-loss and C$_2$H$_2$-loss channels of Q$^{2+}$, IQ$^{2+}$, and Np$^{2+}$ are reversed between O$^{+}$ and O$^{6+}$ impact, indicating that internal excitation plays a key role in the multistep fragmentation of dications. However, the overall contribution of HCN-loss is consistently dominant for both PANHs and both projectiles. These trends demonstrated a higher propensity of PANHs to undergo HCN-loss compared with the analogous C$_2$H$_2$-loss from naphthalene under different internal excitations. Although we observed differences in the BRs of all three neutral-loss channels between quinoline and isoquinoline, HCN-loss shows the most pronounced isomer contrast: IQ$^{2+}$ exceeds Q$^{2+}$ by 4.4\% under O$^{+}$ and by 3.3\% under O$^{6+}$. Thus, the positional identity of the nitrogen atom in quinoline and isoquinoline is reflected mainly in the many-body fragmentation of their dications.

Further, molecular dynamics simulation and PES calculations were carried out to explore neutral-loss and fragmentation mechanisms for Q$^{2+}$ and IQ$^{2+}$. Both dications are predicted to isomerize to common seven-membered ring structures, 1-aza-azulene$^{2+}$ and 5-aza-azulene$^{2+}$, prior to the elimination of C$_2$H$_2$ and HCN, respectively. IQ$^{2+}$ also accesses additional pathways via isomerization to 2-aza-azulene$^{2+}$ and 6-aza-azulene$^{2+}$, leading to C$_2$H$_2$ and HCN loss, respectively. The calculated energy barrier for HCN loss was obtained to be the lowest among the pathways considered, making this channel energetically favored relative to the others, that is consistent with our experimental observations. Moreover, we identified HCNH$^{+}$/C$_8$H$_4^{+}$ (H), HCNH$^{+}$/C$_6$H$_3^{+}$ (C$_2$H$_2$), and C$_3$H$_3^{+}$/C$_5$H$_3^{+}$ (HCN) as the strongest fragmentation channels of Q$^{2+}$ and IQ$^{2+}$ following the respective neutral emissions, and we investigated the corresponding PES in detail. We further evidenced that each of these dominant channels exhibits a delayed fragmentation along with the dominant prompt fragmentation, implying the metastability of the corresponding intermediate species C$_9$H$_6$N$^{2+}$, C$_7$H$_5$N$^{2+}$, and C$_8$H$_6^{2+}$.

Finally, we highlight that the dominant decay products observed in the present work, HCN and HCNH$^{+}$, are abundant in Titan’s atmosphere. The comparison of fragmentation yields and the likelihood of nitrogen-unit elimination suggests that PANHs are less likely to survive in harsh astrophysical environments than their pure-hydrocarbon PAHs counterparts. Our observations indicate the possible role of PANHs in nitrogen containing environments of Titan and even asteroids. With this work, we propose further systematic studies of both dissociative as well as associative processes to comprehend the role of PANHs in the equilibrium composition of Titan’s hydrocarbon-nitrile rich atmosphere.
 
\section*{Supplementary material}
The supplementary material provides the full coincidence map for all three molecules and both projectiles. It contains complete PES for C$_2$H$_2$ and HCN neutral-loss channels of Q$^{2+}$. Fragmentation via additional pathway of IQ is also given. At the end, the graphs and fitting to deduce decay constant of delayed fragmentation channels are described. 

\section*{ACKNOWLEDGMENTS}
This research was supported by funding from the French government, managed by the National Research Agency (ANR), under the France 2030 program, reference ANR-23-EXES-0001. This article is based upon work from COST Action CA21126 -- Carbon molecular nanostructures in space (NanoSpace), supported by COST (European Cooperation in Science and Technology). We acknowledge the generous allocation of computer time at the Centro de Computaci\'on Cient\'{\i}fica at the Universidad Aut\'onoma de Madrid (CCC-UAM). This work was partially supported by the MICINN - Spanish Ministry of Science and Innovation -- Project PID2022-138470NB-I00 funded by {MCIN/AEI/10.13039 /501100011033}, and the `Mar\'ia de Maeztu' (CEX2023-001316-M) Program for Centers of Excellence in R\&D. We also acknowledge the CIMAP and GANIL technical staffs for their support and special thanks to Claire Feierstein for the smooth running of the facility.

\section*{AUTHOR DECLARATIONS}
\subsection*{Conflict of Interest}
The authors have no conflicts to disclose.
\section*{Author Contributions}
\textbf{Sumit Srivastav}: Conceptualization (lead); Data Curation (lead); Formal Analysis (lead); Investigation (equal); Writing/Original Draft Preparation (lead); Writing/Review \& Editing (equal), Validation (equal); \textbf{Sylvain Maclot}: Investigation (equal); Funding Acquisition (lead); Writing/Review \& Editing (equal); Resources (equal); Validation (equal). \textbf{Alicja Domaracka}: Investigation (equal); Writing/Review \& Editing (equal); Validation (equal). \textbf{Sergio D\'{\i}az-Tendero}: Writing/Review \& Editing (equal); Investigation (equal), Resources (equal) ; Validation (equal). \textbf{Patrick Rousseau}: Investigation (equal); Writing/Review \& Editing (equal), Validation (equal), Supervision (lead).
\section*{DATA AVAILABILITY}
The data that support the findings of this study are available from the corresponding author upon reasonable request.

\section*{REFERENCES}
\bibliography{Q-IQ-Np_v2}

\end{document}